# Hybrid FPMS: A New Fairness Protocol Management Scheme for Community Wireless Mesh Networks

Chathuranga H. Widanapathirana[*]  Y. Ahmet Şekercioğlu[†]  Bok-Min Goi[‡]

**Abstract**

Node cooperation during packet forwarding operations is critically important for fair resource utilization in Community Wireless Mesh Networks (CoWMNs). In a CoWMN, node cooperation is achieved by using fairness protocols specifically designed to detect and isolate malicious nodes, discourage unfair behavior, and encourage node participation in forwarding packets. In general, these protocols can be split into two groups: Incentive-based ones, which are managed centrally, and use credit allocation schemes. In contrast, reputation-based protocols that are decentralized, and rely on information exchange among neighboring nodes.

Centrally managed protocols inevitably suffer from scalability problems. The decentralized, reputation-based protocols lacks in detection capability, suffer from false detections and error propagation compared to the centralized, incentive-based protocols.

In this study, we present a new fairness protocol management scheme, called Hybrid FPMS that captures the superior detection capability of incentive-based fairness protocols without the scalability problems inherently expected from a centralized management scheme as a network's size and density grows.

Simulation results show that Hybrid FPMS is more efficient than the current centralized approach and significantly reduces the network delays and overhead.

## 1  Introduction

Wireless Mesh Networks (WMNs) are becoming a leading contender for providing network access in residential communities, specially when the well-established communication infrastructure does not exist [1–3]. WMNs are easy to deploy, have broadband capabilities and can cover large geographic areas without overbearing infrastructure costs. They do not require sophisticated operational expertise because of their ad-hoc nature and adaptability of widely available off-the-shelf hardware [4–6]. The specific application of WMNs in a user community where users are independent of one another is called Community Wireless Mesh Networks (CoWMNs).

In a CoWMN, each node is installed and operated by end users to form the network's infrastructure. Consequently, all network operations are performed at the user nodes. In an ordinary WMN where every node is owned by a single operator, it is possible to deploy mechanisms to maintain tight cooperation among the nodes. However, a CoWMN does not have a central authority and entirely depends on cooperative interaction of the nodes to make end-to-end communications successful [7]. Since the terms of collaboration are always user centric in CoWMNs, traditional network cooperative models are not directly applicable [8].

It is certain that, for a CoWMN having a reasonably large number of users, a percentage of nodes inevitably "misbehave" and become uncooperative, selfish, or malicious [9]. Selfish nodes enjoy network services but refuse to forward others' packets, invalidating the basic collaboration premise of CoWMNs. Malicious ones adopt cheating strategies to hide their unfair behavior.

[*]Department of Electrical and Computer Systems Engineering, Monash University, Melbourne, Australia
[†]Department of Electrical and Computer Systems Engineering, Monash University, Melbourne, Australia
[‡]Department of Electrical and Electronic Engineering, Universiti Tunku Abdul Rahman, Kuala Lumpur, Malaysia

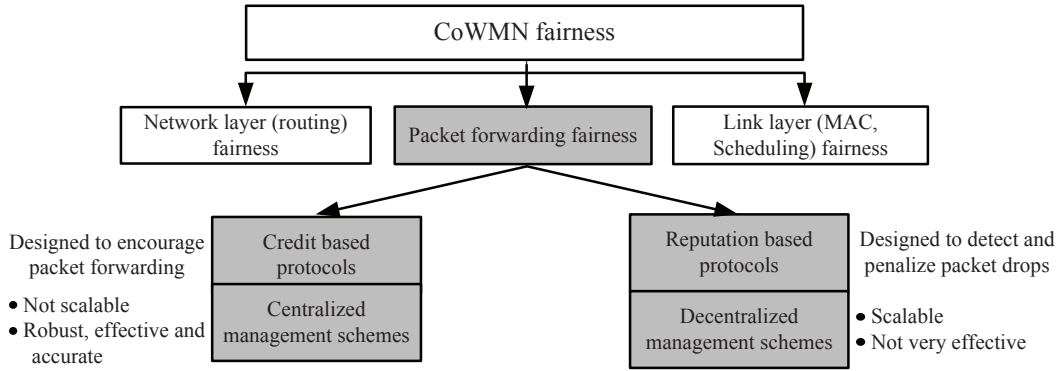

Figure 1: Fair operation of CoWMNs relies on careful design of a number of protocols. In this paper, we focus only on packet forwarding fairness issues which are highlighted in grey.

Properly designed mechanisms are needed to detect, isolate, and punish the selfish and malicious nodes to maintain the fairness.

In all WMNs, the node misbehavior can be caused by unfair behavior of routing protocols or link layer media access control-MAC operations [10, 11]. However, in addition to these common fairness issues of the WMNs, the CoWMNs should also consider the packet forwarding fairness of each node (Figure 1). So far, research community has mainly focused on fair routing and MAC protocol issues within WMN environments. Unfortunately, packet forwarding fairness, albeit its importance to CoWMNs, has been neglected. This study only focuses on the fairness in the context of packet forwarding operations.

In an earlier study, to fill this gap, Widanapathirana et al. [12] have published a study to report their work on packet forwarding fairness in CoWMNs. The algorithm (MPIFA - Modified Protocol Independent Fairness Algorithm) presented in the paper worked well for small to moderately sized CoWMNs, but in larger scale, its performance was poor due to the centralized nature of the management scheme (packet forwarding protocols, and efficiencies of their management schemes are discussed in Section 2 in detail). In this paper, a new fairness protocol management scheme (Hybrid FPMS) capable of eliminating the scalability problems of a centrally managed fairness protocols is presented. In this study, the fairness of the CoWMN is enforced using the MPIFA algorithm. However, instead of centrally deploying MPIFA, we use Hybrid FPMS scheme to combine the original central algorithm with a decentralized management.

The paper is structured as the following. First, an overview of the current fairness protocols in the context of packet forwarding is presented. Then a detailed description of the Hybrid FPMS and the implementation details are discussed. Finally, we analyze the performance through simulations and demonstrate that the fairness operation of the network has shorter delays, smaller overheads, requires less processing resources when Hybrid FPMS is used. By employing an incentive-based fairness scheme we also have the added advantage of improved malicious detection, reduced false detections and detection error propagations [9, 13–15].

## 2 Packet forwarding fairness protocols and management

In this section, we summarize the general approaches to packet forwarding fairness protocols and their associated management schemes. Before proceeding further, it is prudent to differentiate between a fairness protocol and its management scheme. A fairness protocol is an algorithm designed to detect malicious nodes and compel them to cooperate in CoWMN environment. A management scheme's task, on the other hand, involves how a fairness algorithm is deployed or implemented in a network. It should be noted that, the deployment mechanism of a fairness protocol has a significant impact on the effectiveness and efficiency of the protocol and overall operational performance of a CoWMN itself.

Since the literature lacks fairness schemes tailored especially for the CoWMNs, a sensible approach is to look into proposed approaches for Mobile Ad Hoc Networks (MANETs) and WMNs. MANETs and WMNs have close similarities to CoWMNs and number of fairness protocols, and management schemes can be used as starting points for CoWMNs [9, 16–19].

There are two classes of cooperation mechanisms: reputation-based schemes and incentive-based schemes [15, 20]. We briefly describe these in the following sections.

## 2.1 Reputation-based schemes

A number of schemes use "reputation" of a node to measure and mitigate the selfish node behavior. Reputation is a node rating concept inspired by online auctioning systems [18, 21] that provide a means of obtaining a quality rating of a participant of transaction. Reputation-based protocols concentrate mostly on detection and punishment of malicious nodes in a network instead of rewarding and stimulating cooperative behavior. Leading applications of this approach are Watchdog and Pathrater [17], and CONFIDANT [18].

Each node's reputation value reflects its behavior, and is obtained either by direct observations or through the reputation messages generated by the other nodes in the network. A node's reputation value is calculated and stored by its neighbors who monitor how the node behaves. As such, the fairness mechanisms needed should take care of the calculation and update of reputation values, detection of misbehavior, and reaction to uncooperative behavior.

Reputation-based algorithms inherently use decentralized protocol management mechanisms, and have a higher erroneous detection ratio and inferior detection accuracy on identification of malicious nodes than incentive-based schemes [9, 14, 15]. Another drawback is the possibility of "false accusation propagation" [22, 23] throughout the network due to lack of coordination. Furthermore, since every node equally carries the burden of fairness protocol, decentralized management demands every user to allocate memory and processing power regardless of resource availability.

However, in decentralized schemes, the propagation latencies are small and independent of the network size due to the constant one or two hop paths. The amount of processing required for fairness decisions and amount of information exchanged are also kept consistence regardless of network size.

## 2.2 Incentive-based schemes

Incentive-based techniques use "micro-payments" to motivate the nodes to cooperate. A node receives credit if it serves the network and pays back a price when it uses the network. Leading applications of this approach are MPIFA [12], PIFA [24], Packet Purse Model (PPM) with Packet Trade Model (PTM) [19] and SPRITE [9]. Incentive-based systems are considered simpler algorithms and require little computational resources at each node compared to reputation-based schemes because nodes are required only to keep track of packet exchanges and credit limits while the burden of fairness decision-making is shifted elsewhere. Even though an incentive-based system is capable encouraging fairness in the network, they do not have a mechanism for preventing the malicious behavior of nodes.

Most of the incentive-based systems are implemented as centralized systems where a Central Management Server (CMS) keeps track of all the credit transactions in the network. CMS is tasked with gathering information from every node to collectively process and disseminate the fairness decisions. However, having a centralized management scheme in a CoWMN creates two major challenges to the efficient scalability of the protocol:

- The first challenge is the delay caused by having to transmit update messages from every node to a central server and back. Two types of delays occur during such transfer, one being *Link Propagation Delay* and other being *Node Processing Delay*. Each update has to be routed through multiple nodes, increasing both aggregated propagation delay and node processing delays per transfer.

- The second challenge is the additional overhead added to the network traffic by fairness updates serving large amount of nodes, which consume valuable bandwidth available for the users.

From the comparison of management schemes above, it is obvious that either one of them, on its own, is insufficient to tackle the challenges posed by malicious nodes. Decentralized management performs efficiently with network growth, but the reputation-based fairness protocols lags behind the incentive-based protocols when detection accuracy, robustness and resource requirements are considered. Centralized management lacks the ability scale efficiently when a network grows, even though the incentive-based protocols perform well in small CoWMNs. Hence, our hybrid approach combines the characteristics of above two management techniques to create a more efficient and robust fairness mechanism for CoWMN.

## 2.3 Modified Protocol Independent Fairness Algorithm (MPIFA)

Protocol Independent Fairness Algorithm (PIFA) was originally introduced by Yoo et al. [24]. This algorithm implements node fairness in (MANETs) and has the capability of detecting and removing malicious nodes reside within MANETs. Widanapathirana et al. [12], extended the PIFA algorithm to CoWMNs by creating the *Modified Protocol Independent Fairness Algorithm (MPIFA)*. MPIFA is an incentive-based protocol, but also encompasses a reputation matrix unlike other such algorithms. As the name suggests, MPIFA operates independent of the network layer utilizing any available routing scheme, thus flexible enough to be used in any CoWMN. A brief summary of operation of MPIFA is given here to introduce the algorithm.

The MPIFA is managed centrally with a Central Management Server (CMS) which has a Number of Alleged Manipulations (NAM) table (reputation matrix) and Credit Database (CDB). When deployed, each user node keeps records of packet transactions with every other node and are sent then to the CMS in regular intervals. The records from the nodes contain

(i) $I$, the input packets from the neighbor,

(ii) $O$, the output packets to the neighbor,

(iii) $S$, the generated packets sent to the neighbor,

(iv) $T$, terminated packets sent from the neighbor,

(v) $OFN$, packets originated from the neighbor itself among input packets from the neighbor pertaining to every neighbor.

Given, $a$ and $b$ are neighbors and assuming no collusion between nodes, the CMS carries out three tests to identify the malicious behavior towards the fairness mechanism as follows:

$$O_{a,b} = I_{b,a} \tag{1a}$$
$$S_{a,b} = OFN_{b,a} \tag{1b}$$
$$\sum I_{a,\forall b} - \sum T_{a,\forall b} = \sum O_{a,\forall b} - \sum S_{a,\forall b} = F_a \text{ where } F \text{ is total packets forwarded} \tag{1c}$$
$$\tag{1d}$$

When a test is failed, since the CMS can not unambiguously determine the malicious node, the NAM table is updated by a penalty of $X$ for both $a$ and $b$

$$NAM_{a,b} = NAM_{a,b} + X \text{ and } NAM_{b,a} = NAM_{b,a} + X. \tag{2}$$

Assuming malicious node highly likely to repeat the attempts to deceive others, other nodes (*i*) NAMs that have been accumulated in connection with nodes *a* and *b* before are reduced by a factor of *Y*

$$NAM_{i,a} = NAM_{i,a}/Y \text{ and } NAM_{i,b} = NAM_{i,b}/Y. \tag{3}$$

Once the $\sum NAM_i$ of the $i^{th}$ node exceeds a threshold, the node is blacklisted.

The CDB keeps track of credit by updating $CDB_a$ after every update using

$$CDB_ia = CDB_a + F_a \times \beta - \sum S_a \times \delta \tag{4}$$

where $\beta$ is the forwarding reward and $\delta$ is the generation cost. The fairness is encouraged by only allowing the nodes with a minimum credit limit to generate or terminate traffic. Due to the space constraints further details and performance of the MPIFA can be found in [12].

## 3 Hybrid Fairness Protocol Management Scheme

### 3.1 Hybrid FPMS concept

Here, we present a new protocol management scheme called, *Hybrid Fairness Protocol Management Scheme (Hybrid FPMS)*. This scheme uses a concept of distributed virtual zones to deploy a centralized fairness protocol with a degree of localized management. Hybrid FPMS is designed to operate independent of network stack and utilizes the available protocols for node-to-node communications. In this paper, MPIFA is used as the underlying fairness algorithm to demonstrate the operation of Hybrid FPMS. However, with simple modifications, Hybrid FPMS can also be used with other centralized algorithms as long as they operate independent of routing protocols.

Hybrid FPMS can be divided into four operational phases as shown in Figure 2 for better understanding of the overall concept. Hybrid FPMS creates network zones, which are collections of nodes virtually grouped together within the context of packet forwarding fairness. Instead of considering whole CoWMN as a single domain, MPIFA is implemented to each zone independently, isolating the protocol operations within the respective zone. Since MPIFA requires a CMS, we introduce a functionally similar Zone Management Server (ZMS) to each zone. ZMS receives the fairness updates and makes the fairness management decisions for the particular zone. With this approach, we remove the fairness management functionality from the CMS, but allow for the CMS to function as the administrative hub for overall CoWMN. Mechanisms are implemented to establish limited communication between ZMSs and CMS in order to coordinate the network.

### 3.2 Operational details of Hybrid FPMS

#### 3.2.1 Creating virtual zones

The first step of the Hybrid FPMS is to create virtual zones. Zoning is performed as part of the network initiation process and any new subscriber is assigned to a zone following the same zoning criteria. When the CoWMN landscape evolves significantly from the state of initiation, a rezoning of the network can be performed.

Zoning is handled by the CMS using the subscription information it retains about every node. A special consideration is given to the node's relative geographical and topological proximity. The number of nodes in each zone has a significant impact on the overall performance of the system. In a previous study, it has shown that MPIFA provides a good balance between performance and efficiency in a CoWMN with approximately 40 nodes [12] using simulations. Too many nodes in a zone can diminish the intended efficiency and with few nodes per zone. In contrast, the detection accuracy can be effected because of the smaller sample size.

One approach is to create zones using GPS data to isolate nodes within close proximity i.e. geographical approximation. With this approach, a topological approximation can be assumed if omnidirectional nodes are used because of the nature of the radio networks. Even though this zone assignment is computationally simple, it requires additional cost of GPS devices to be incorporated into the network. Furthermore, this approach does not result in the optimum zone assignments when nodes use directional antennas. Hence, we propose a heuristic-based topological zoning algorithm.

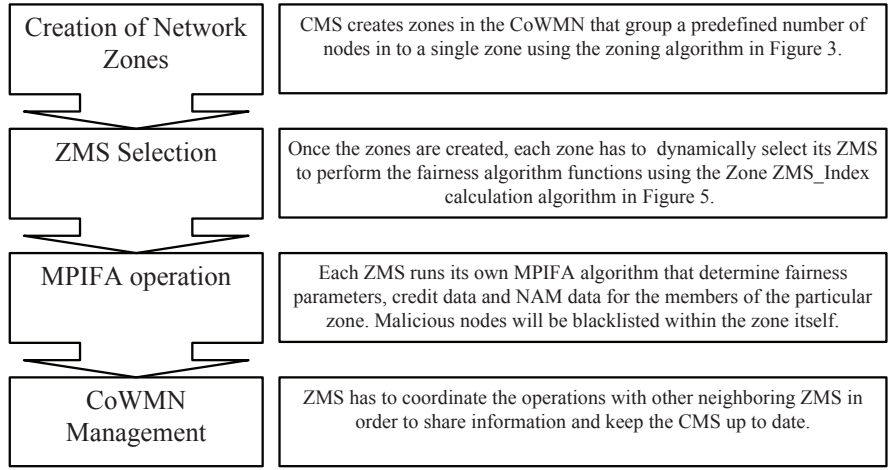

Figure 2: Main operational phases of Hybrid FPMS

### 3.2.2 Zoning algorithm

We assume each node has provided CMS with data such as the type of node infrastructure, an indication of processing power, available resources, and neighboring nodes. CMS generates a topological map of the network during network initiation using the neighbor information. This map is used as a reference to dynamically grow zones using the zoning algorithm in the flow chart in Figure 3.

**Stage 1** Select a node with the least number of neighbors (border node) from available nodes listed. This node is used to grow the zone around it. By selecting a node with least neighbors to start the zone we ensure zones are started from a border region first and no isolated pockets of nodes are left without a zone.

**Stage 2** Add the neighbor nodes of the selected node to the zone list. Remove them from the available list. After each node is added, zone list is checked to see node limit per zone ($\approx$(Total Subscriber Base / Nodes per Zone)) is reached.

**Stage 3** Once all the first-tier neighbors(of the selected node) are added to the zone list, newly added nodes has to be rearranged in the list from node with the smallest number of neighbors to node with the highest number of neighbors.

**Assumption** - nodes that are situated at the edges of the CoWMN have a smaller number of neighbors in comparison to nodes that are located in the middle. This assumption is accurate because nodes at the edges of network only have neighbors in less than 360 degrees (mostly less than 180 degrees) while nodes that are located in the middle of the network are connected to nodes all around (360 degrees).

Stage 3 is followed to make sure that all the nodes in the edges of the network are added to the zone. If we do not follow this step and add the second tier starting from random first-tier node, some nodes in the edges can be left without a zone.

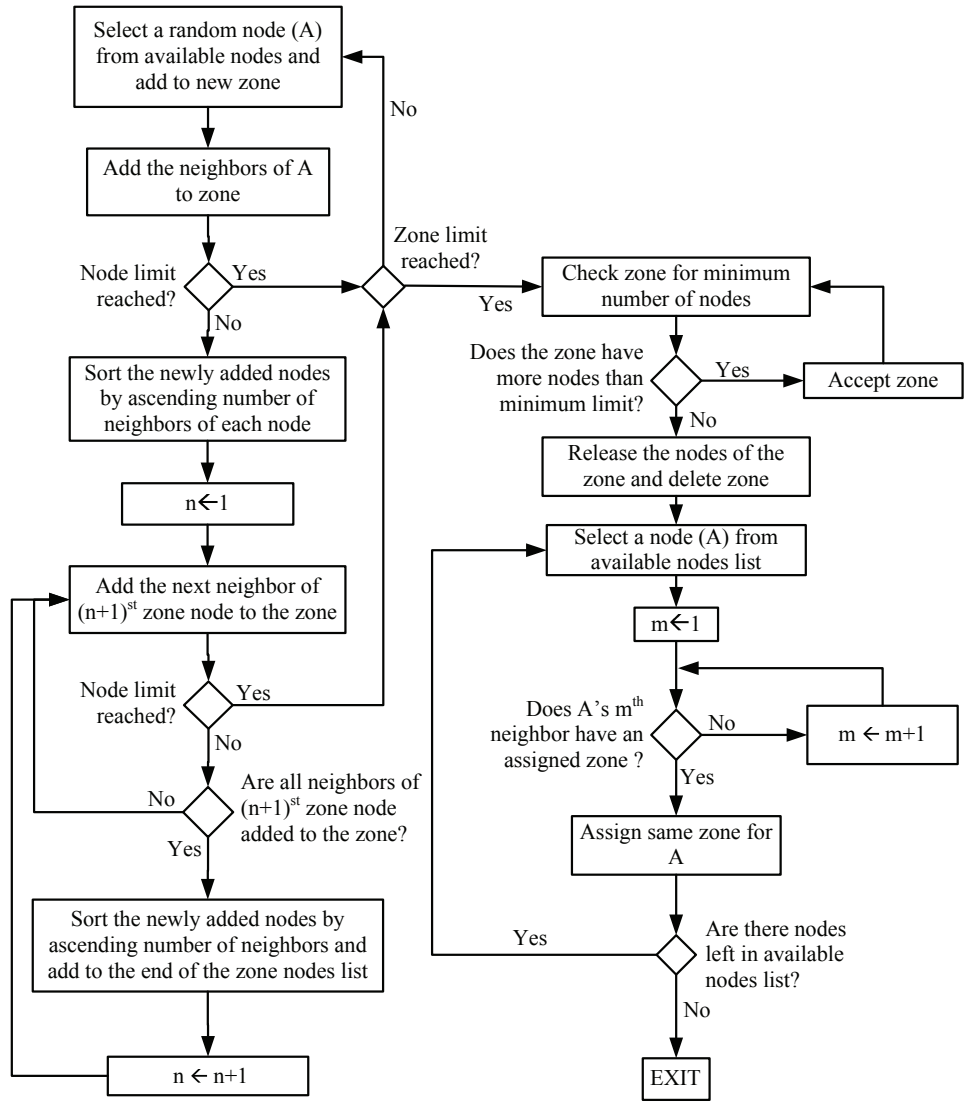

Figure 3: Flowchart of the zoning algorithm.

**Stage 4** We start to add from the 1st neighbor of the 2nd node in the zone list. Every time a node is added, we perform a check to see zone has reached its node limit. If not, we keep on adding neighbor nodes until 2nd node in the list runs out of neighbors. When nodes from second tier onwards are added, they are first added to a temporary buffer to rearrange them in ascending order of the number of neighbors. Subsequently, they are appended to the zone list.

**Stage 5** Once 2nd node in the list runs out of neighbors we move to $3^{rd}$ node in the zone list to add the neighbors of that particular node. We follow the same step until the zone node limit is reached or all the CoWMN nodes are assigned a zone. We do so because each iteration adds a node to the zone list and keeps on extending the list so that algorithm can always find a new node to add its neighbors to the zone.

**Stage 6** Once all the zones are created, the algorithm checks if the zone has a minimum number of nodes specified. If it does not have the minimum number required for a zone, the nodes in the zone are released again to the available node list, and the zone is deleted. This is done to avoid creating small zones that affect the effectiveness of MPIFA algorithm. Following step is performed to add these nodes to the available best possible zone.

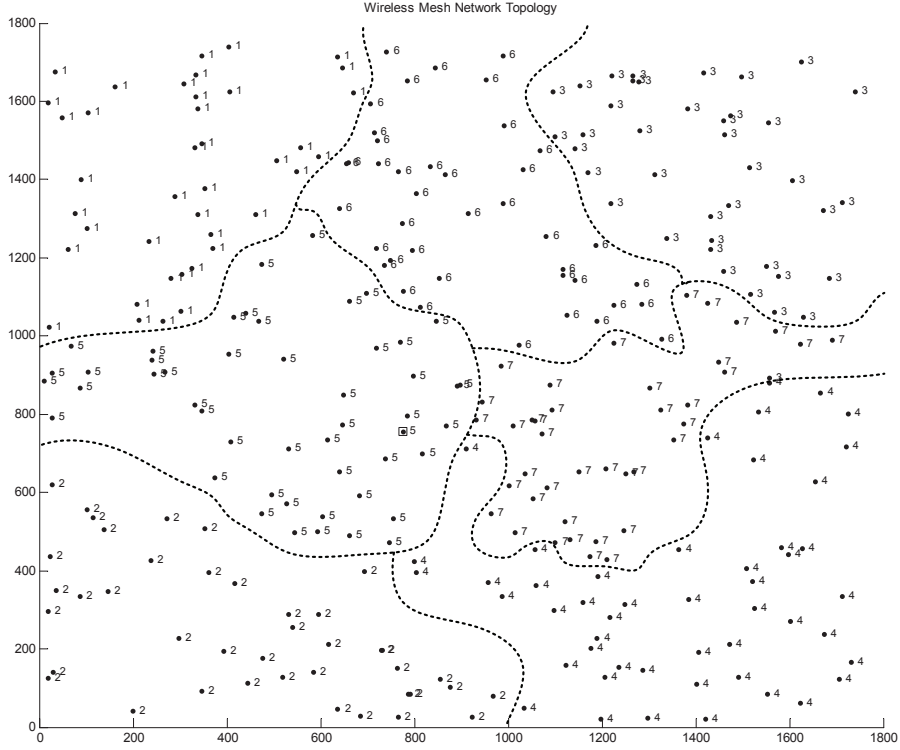

Figure 4: Zones created using the zoning algorithm.

**Stage 7** First neighbor of each node in the available node list is checked to see whether it has a zone assigned to it and if so this particular node is added to the same zone. If the first neighbor does not have a zone assigned, algorithm checks all the neighbors until it finds a neighbor node with a zone assigned. This process is done to all the remaining un-zoned nodes until all the nodes are assigned a zone.

Figure 4 illustrates the simulated results of the zoning algorithm in a CoWMN with 300 nodes and 0.0001 (nodes per $m^2 \Longleftrightarrow$ 100 nodes per $km^2$ ) node density. Network is zoned into 7 zones with approximately 40 nodes in each zone. It shows that our zoning algorithm sufficiently fulfills the criteria of zoning.

#### 3.2.3 Selecting Zone Management Server (ZMS)

Once the zones are created, next step is to assign a Zone Management Server (ZMS) to each zone. CMS handles the process of ZMS assignment. CMS uses the information sent by each node on available resources to calculate the optimum ZMS node. Criteria for selecting the ZMS are twofold to guarantee the best node capable of maximizing the zone efficiency. First, ZMS has to possess enough resources to process and store all the fairness information without disrupting the functions and performance of the user. Second, ZMS node should be easily accessible with minimum latency by each node in the zone.

To identify the optimum ZMS, we propose a node *ZMS_index*, which indicates the nodes suitability to be the ZMS. We use another parameter called *Node_Delay_Index*, which represents the maximum delay induced by the node in processing, queuing, and transmitting a fairness update message. We assume that Node_Delay_Index value is known for each node. This is derived from the resource updates sent to CMS. However, calculating the relationship between available node resources and delays is out of the scope of this publication. During ZMS selection, we also assume that link propagation delay through the medium is negligible in comparison to processing delay.

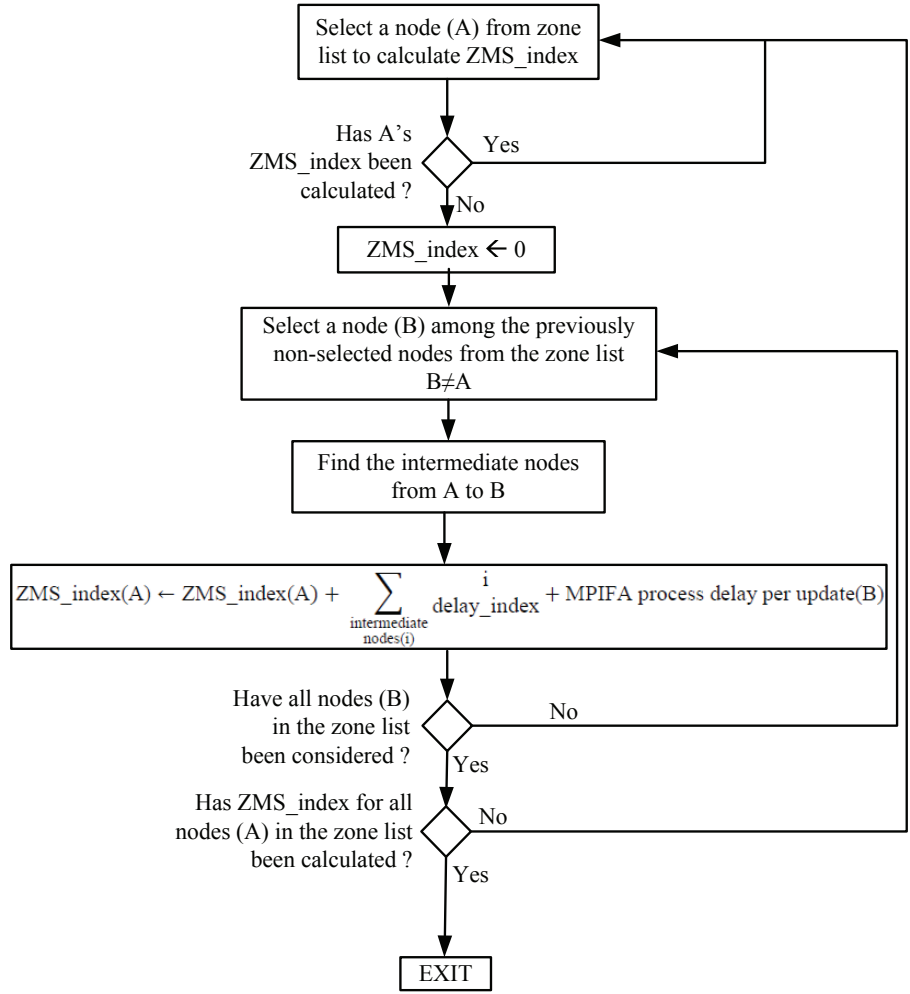

Figure 5: Zone ZMS_Index calculation algorithm.

ZMS_Index value is calculated using the Node_Delay_Index value as shown in Figure 5. ZMS_index for a node is calculated by finding the sum of the path Node_Delay_Indexes from all the other node in the zone to this particular node. This technique helps to identify the easily approachable nodes of the zone. Apart from the path Node_Delay_Index, processing delay incurred at node A to process the fairness message sent by node B also has to be considered.

Once all the ZMS_Indexes for the zone are calculated, node with the lowest ZMS_Index value (most easily accessible) is selected as the zone ZMS. Subsequently, CMS initializes the Credit Database (CDB); NAM matrix required by the MPIFA algorithm and informs every node in the zone about the identity of ZMS node. From this point onwards, nodes communicate only with ZMS.

### 3.2.4 Integration of fairness algorithm

The initialized virtual zones provide a perfect platform with suitable size and configuration to deploy MPIFA. MPIFA is operated in relative independence within each zone to manage the fairness. In the Figure 6, circle shows few neighbors of the node X (of zone 7). When data communications proceeds, the node X records transactions with node in zone 4, 5, 6 and 7. Because MPIFA requires all packet exchanges done by a node to be recorded, border nodes that had communications with adjacent zones update ZMSs of those zones as well. Since nodes do not have knowledge of the location of the neighbor zone ZMS, each node with neighbors in the neighboring zones has to update those neighbors with location of its own ZMS during network initiation. With this knowledge, if a node has a record of inter-zone packet exchange, a fairness update is sent to ZMS of that particular zone.

### 3.2.5 Zone coordination and CoWMN management

MPIFA requires few rounds of ZMS updates before the accurate detection of malicious nodes. Once each ZMS starts detecting malicious behavior, it updates its servicing nodes about the detected malicious nodes and black-list them from the zones' normal network operations. By notifying all the nodes, ZMS makes sure that active nodes drop all the packets originated by and destined to the blacklisted nodes. These nodes are also avoided in the routing.

The ZMS also instructs the boundary nodes to propagate the blacklisting information to the nodes neighboring the particular zone. By doing so, all the nodes that perform inter-zone communications are aware of the malicious nodes in neighbor zones and prevent malicious nodes in the borders from using adjacent zone to transmit their packets.

Use of ZMS avoids the unnecessary communication between nodes and CMS. By doing so, the network loses the transparency towards CMS. However, the prudent way is to keep track of credit fluctuations and malicious behavior in the CoWMN by administrators to prevent blacklisted nodes from re-subscribing. This information also can be used to legal purposes and billing purposes and can be valuable for future developments. Because of that, each ZMS updates the CMS with credit variations of member nodes and the malicious behavior statistics. This transfer occurs during periods of lower network utilization and load. Since the amount of data to be ex-changed between management servers is significantly large, it is important to find suitable time periods to transfer these updates to keep them from affecting the network performance.

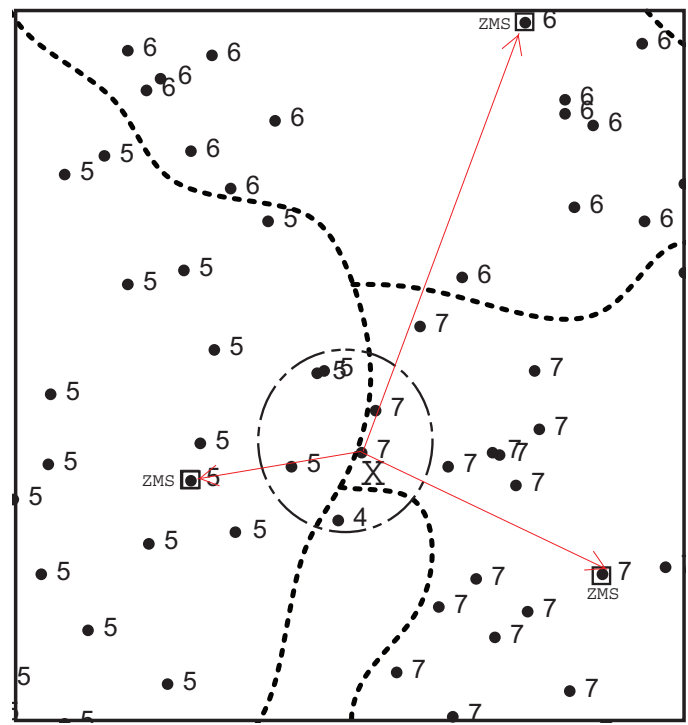

Figure 6: Inter-zone communication of border nodes.

## 4 Performance evaluations of Hybrid FPMS

Performance evaluation was done using a MATLAB based mesh network simulator. In [12], authors demonstrated the superior ability of MPIFA to detect malicious nodes. In our performance evaluation, we focus on comparing MPIFAs' centralized management scheme against the Hybrid FPMS to demonstrate its greater efficiency and robustness in large-scale CoWMNs.

We keep the malicious node ratio at 30% to represent a worse-case scenario, link failure probability at 10% for this analysis. Malicious behavior of a node is simulated by dropping packets being forwarded through such a node with a given probability (Malicious behavior probability of 80% for this analysis). Management server update delays, malicious nodes detection time, and fairness traffic overhead ratio has the biggest impact on fairness protocol efficiency and overall network performance. We focus our evaluation on these three factors over various network sizes and node densities.

### 4.1 Removal of malicious nodes

Figure 7(a) shows a simulated CoWMN of 200 nodes, randomly distributed in $1500m \times 1500m$ area. Network is divided to 4 virtual zones of 50 nodes each by CMS during initiation. 30% of nodes are simulated to behave maliciously. With this simulation setup, we run MPIFA deployed using Hybrid FPMS to verify the feasibility of our concept. Figure 7(b) shows a snapshot of the network short time after the network is operational and from that, we can observe that malicious nodes have been removed from the network as expected. This demonstrates the feasibility of our approach to deploy a centralized algorithm, partially distributed in a CoWMN and still perform the fairness function.

If a large portion of CoWMN nodes is malicious, MPIFA/Hybrid FPMS can create a partially connected CoWMN. Since removal of the malicious nodes results in fewer alternative paths, a performance tradeoff between reliability of CoWMN and fairness is expected when the malicious node ratio reaches higher proportions even though it is highly unlikely in a practical scenario. This complex relationship depends on a multitude of factors such as node density, node range and routing protocols.

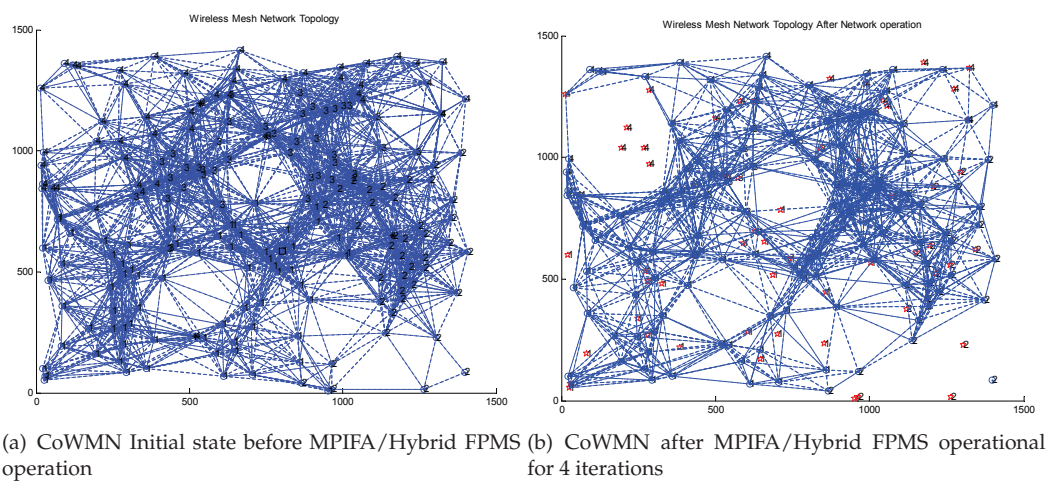

(a) CoWMN Initial state before MPIFA/Hybrid FPMS operation (b) CoWMN after MPIFA/Hybrid FPMS operational for 4 iterations

Figure 7: Removal of malicious nodes with MPIFA/Hybrid FPMS.

## 4.2 Network coverage scalability

### 4.2.1 CMS/ZMS update delay

When a CoWMN is deployed, the subscriber base increases over time expanding the network geographically. The ability to grow without having to add new base stations to cover extra areas is one of the positive traits of a CoWMN that makes it flexible and economical. However, when MPIFA (or any centralized fairness protocol) requires a CMS, new nodes joining the network from outer reaches create protocol efficiency degradations.

The results in Figure 8 illustrate the effects of network scale has on the average central server update message delay in CoWMN. We analyzed this effect in 3 different node density levels, 0.00005, 0.0001, 0.00015 (nodes per $m^2$).

Simulated results show that average delay of the CMS update increase with the number of nodes (or scale of the WMN) for centralized MPIFA in all node densities. The average distance a packet traverse and the average intermediate hops increase when the network expands. This creates a higher total link propagation delay and also a higher total node processing delay which adds up to a longer CMS update delay. Since every node has to send its update messages to CMS before fairness decisions are made, network has to enforce strict node limits to avoid additional delays ensued by network growth.

However, with Hybrid FPMS, the average central server update delay shows significant improvements. When the network size increases as in Figure 8, the average ZMS update delay remains constant with only minor fluctuations. In Hybrid FPMS, each node updates its own ZMS instead of CoWMN CMS. Each node only has few hops to the ZMS regardless of its position in the network. By creating a scheme that only requires few hops for an update message; we have eliminated increasing central server update delays in the CoWMN. Since the zones only consist of 40-50 nodes, the average delays are approximately equal to that of centralized network with the similar number of nodes. The simulation shows that each zone acts independently when performing the fairness operations. Percentage improvement in average delay also increases with the number of nodes in CoWMN. With 80 nodes in the network, improvement of delay performance is about 33% while with 200 nodes, we observe a significant improvement of 81.5%. By keeping the delays relatively constant, the CoWMN can be expanded to a larger area with more users without the burden of fairness protocol efficiency degradation.

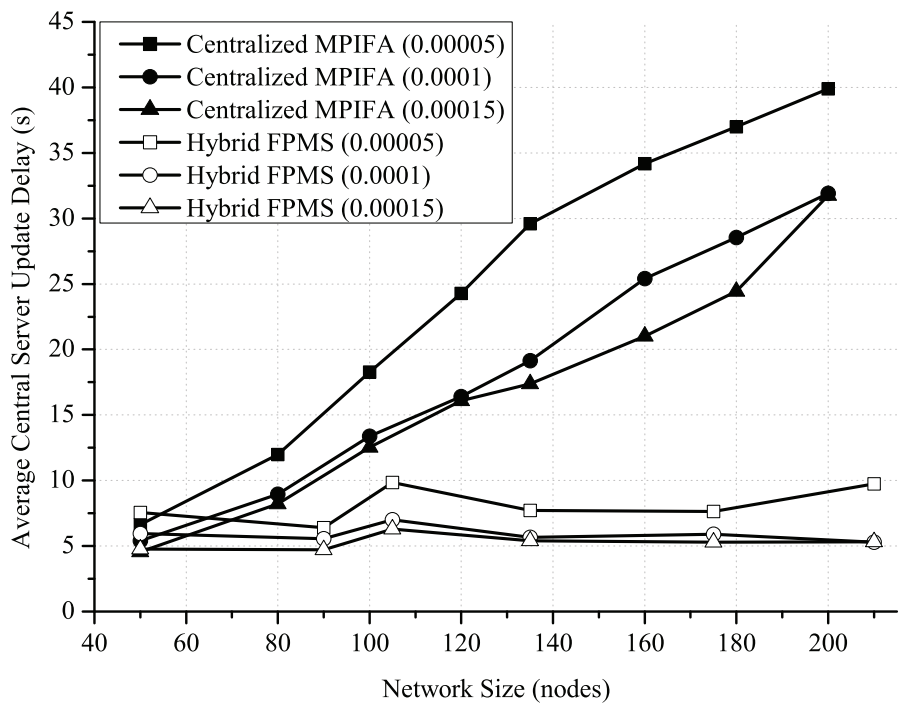

Figure 8: Average central server update delay vs increasing network size.

### 4.2.2 Malicious node detection time

When update delays remains small, ZMS receives enough information to make fairness decisions in a much shorter period. This ensures a faster malicious node detection time. Results in Figure 9 shows the time taken to detect and remove 90% of nodes showing malicious behavior. As explained previously, the average delay to update central server increases with nodes when centralized MPIFA is used. Since every iteration of MPIFA is performed only after all the updates are received, it takes increasingly longer time to detect the malicious nodes when the network grows in scale. Apart from propagation delays, CMS of the network has to analyze, process and calculate information of all the nodes in the CoWMN. Having only one CMS worsen the detection time when the number of nodes supported by CMS increase. Simulated results in Figure 9 shows how detection delays are increased with nodes. Longer time to detect malicious nodes negatively impact overall network performance because of higher packet drop rates [12].

When Hybrid FPMS is used, the average delays are kept at a lower constant value regardless of increasing number of nodes. This translates into significantly improved detection times as reflected from Figure 9. The time taken to detect malicious nodes has improved by 15% for a CoWMN with 80 nodes and improvement is about 75% for a much larger network with 200 nodes. Furthermore, capability of Hybrid FPMS to keep the detection times constant provide a window for users to be cautious during that specific period. When Hybrid FPMS is used, the number of nodes each ZMS has to process is significantly lower than that of a centralized system. This reduces the processing burden on the node ensuing faster detections. These results again reinforce the observations from previous discussion to prove that Hybrid FPMS gives a better scalability to the CoWMN.

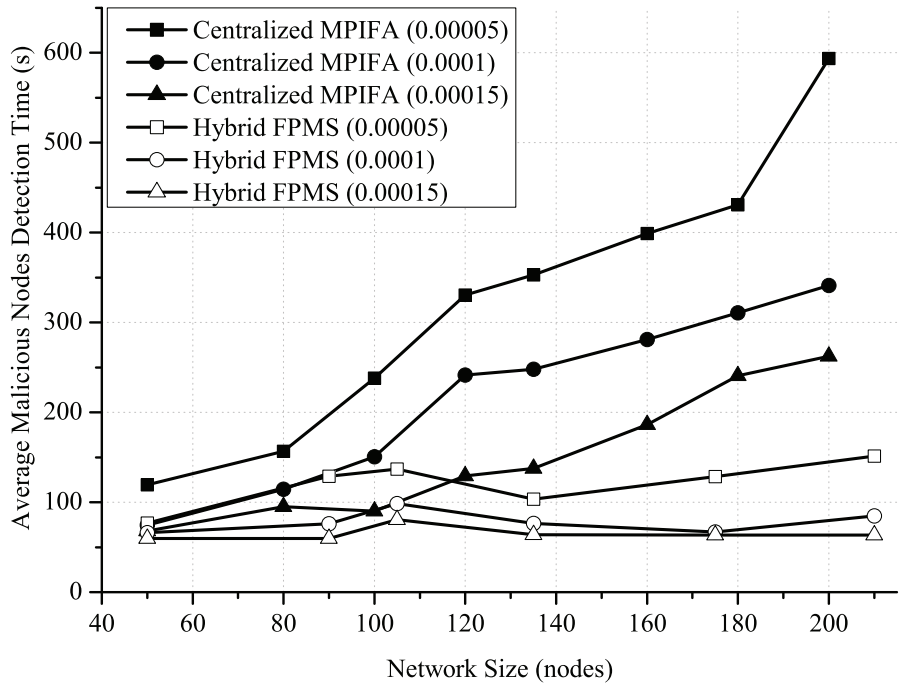

Figure 9: Malicious nodes detection time vs increasing network size.

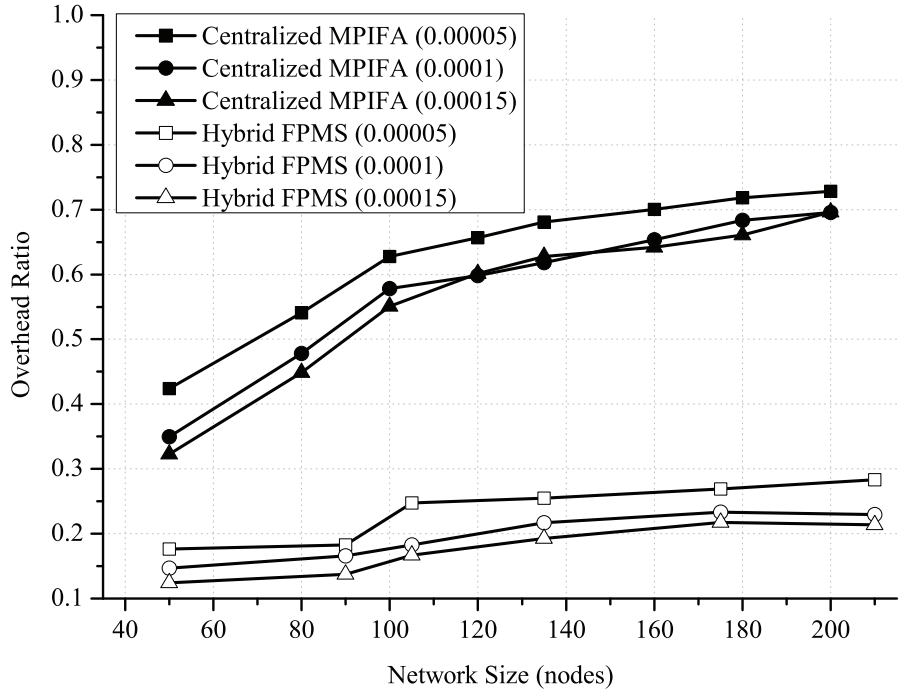

Figure 10: Network overhead ratio vs increasing network size.

### 4.2.3 Network overhead

Updating CMS and ZMS with fairness messages in frequent periods create additional traffic in CoWMN. As authors demonstrated in [12], reducing the update frequency has a negative impact in the MPIFA performance. Hybrid FPMS is designed to reduce the size of messages created by fairness protocols, which result in a significantly lower network overhead ratio between fairness protocol traffic and other traffic.

There are two types of control messages being exchanged between the nodes and Management Server (MS). Nodes send update message to MS containing traffic information pertaining to their neighbors and once the processing is done MS returns information on blacklisted malicious nodes. From the Figure 10, we can observe that when using centralized MPIFA, a very large portion of the network traffic consists of overhead data. With MPIFA, each node sends update packet that contains traffic exchange information with all the neighbors to CMS and CMS send a packet that contains information about nodes' validity and credit warnings. Three factors; number of update messages to the CMS, number of update message from the CMS, and size of the message (message length) increase with the network growth that ultimately result in the increased overhead in the CoWMN.

When Hybrid FPMS is being used, we observe a significant reduction in the network overhead. The nodes sends update messages, some to its own ZMS and few to its neighbor ZMS resulting in overhead. However, the update messages only contain information about exchanges with nodes belong to that particular ZMS. Hence, the packet size is comparatively smaller than that of a centralized algorithm. But, there is a significant difference in the size of the message sent from ZMS to update nodes because each ZMS only services its zone members and message only has to contain information about few nodes. Increase in the number of nodes results in more zones being created but the average number of nodes in a zone remains the same. Because of that, size of messages exchanged both ways kept at a constant level regardless of the network size. As shown in Figure 10, Hybrid MPIFA shows 60% improvement of the overhead ratio with 100 nodes network and 62% improvement with 200 node network against centralized MPIFA.

## 4.3 Network density scalability

CoWMN deployed over an urban or a metropolitan area has new users subscribe within the available coverage area which increases the node density. This affects the efficiency parameters such as the delay and overhead ratio. Simulations were performed to analyze the effects of node density changes on delays and overhead in a fixed area of $1200m \times 1200m$. Figure 11(a), Figure 11(b), and Figure 11(c) show average central server update delay, the malicious nodes detection time and network overhead ratio changes against different node densities respectively. As evident from the graphs, the variation of average server update delay, the malicious nodes detection time and the overhead ratio are more or less identical to changing network coverage (size). Increase of the node densities in a fixed area is a direct result of more nodes or users subscriptions. When number nodes situated within a unit area increase, the distance between two nodes is expected to decrease, and more neighbors are added to every node.

### 4.3.1 ZMS update delay and malicious node detection delay

Central server update delay depends on the delay occur in link propagation, processing in intermediate nodes and processing at the central server. The average distance between two nodes reduces when node density increases unlike in the previous case where network coverage grew with little effect on node distances. However, in multi-hop wireless networks, delay occurs in

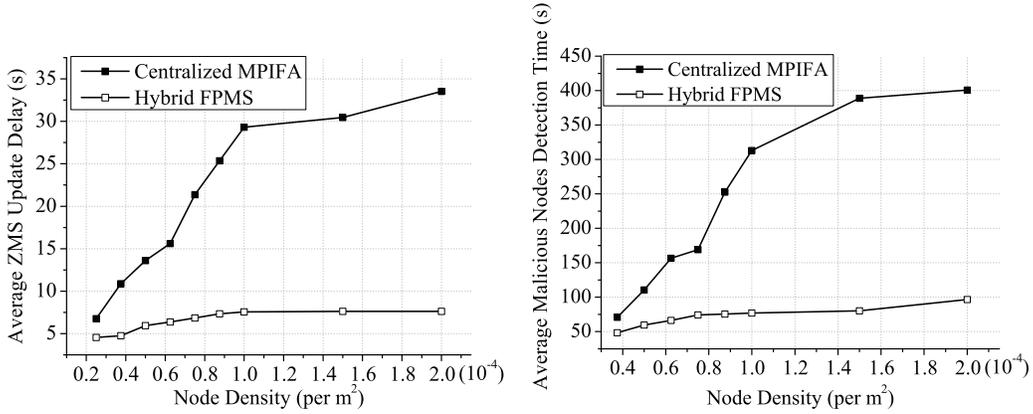

(a) Average central server update delay Vs increasing network density

(b) Malicious node detection delay Vs increasing network density

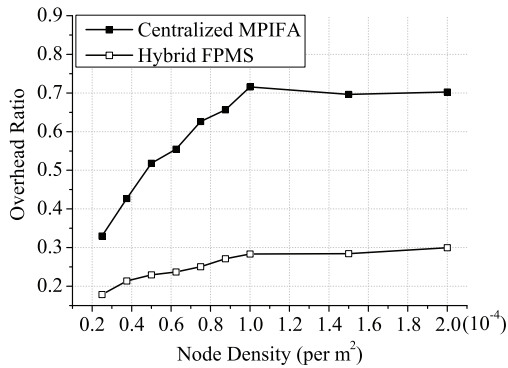

(c) Network overhead ratio Vs increasing network density

Figure 11: The graphs showing the changes in update (Figure (a)), detection (Figure (b)) delays and overhead ratio (Figure (c)) against increasing network density.

node-to-node propagation in the physical medium is significantly smaller in comparison to delay occur due to queuing and processing in intermediate nodes. Because of this reason, the effects of decreasing distance between nodes on the central server update delay in Figure 11(a) is minuscule to be noticed and can be ignored. When that factor is ignored, intermediate processing and queuing delays, node data processing delay at CMS shows identical trends between growth of network size and network densities because both those phenomena ultimately produce more nodes. Even though the congestion characteristics of outer edges are different, congestion surrounding the central server also shows similar trends in both cases. Hence, Hybrid FPMS shows superior delay performance with increasing node densities (Figure 11(a)) owing to the same reasons. Similar delay characteristics in both cases also result in identical detection delays, where as shown in Figure 11(b), Hybrid FPMS improves malicious node detection times for high density networks, just as we observed previously in Figure 9.

### 4.3.2 Network overhead

Effects of increasing number of user nodes due to scale expansion and due to new subscribers in a fixed area are similar because there is no direct correlation between overhead ratio and node density. Since the number of nodes is the only changing factor affecting the network overhead when CoWMN is getting denser, trends in Figure 11(c) and Figure 10 are similar. Hybrid FPMS shows constant low levels of overhead regardless of the densities unlike when a centralized protocol being used.

From these results, we demonstrate that Hybrid FPMS provides significant delay and overhead ratio performance improvement in comparison to centralized MPIFA during both CoWMN network size/coverage expansions and user density increments.

## 5 Conclusion

The objective of the study presented in this paper is to design a scalable packet forwarding fairness management scheme which allows a centralized fairness protocol to be deployed in CoWMN without the inherent performance deterioration associated with the network growth. We have identified that centralized fairness protocols can be extended to operate efficiently and effectively in large scale CoWMN by using a novel approach to protocol management. We introduced Hybrid FPMS, a fairness protocol management scheme that uses centralized MPIFA as the underline fairness protocol, and we demonstrated the implementation procedure of such a scheme in a CoWMN. We created zones to divide the network into more localized segments and used a ZMS to centrally manage the MPIFA protocol independently for each zone.

From the simulated network, we observed that Hybrid FPMS can successfully reduce central server update delays by 33%-81.5% depending on the network size. The faster updates resulted in reduced malicious nodes detection times up to 75%. Hybrid approach has the added advantage of approximately 60% less network overhead, offering productive bandwidth utilization. We also proved that Hybrid FPMS not only can provide improved coverage scalability, but also can provide substantial network density scalability without impacting the efficiency. Furthermore, with the introduction of Hybrid FPMS concept, we offer the flexibility to extend any type of centralized fairness protocol to a large CoWMN without exacerbating the efficiency and throughput performance.